\documentclass[runningheads]{llncs}

\usepackage{eccv}

\usepackage{eccvabbrv}

\usepackage{graphicx}
\usepackage{booktabs}
\usepackage{color}
\usepackage{multirow}
\usepackage{array}
\usepackage{textgreek}

\usepackage[accsupp]{axessibility}  

\usepackage[pagebackref,breaklinks,colorlinks]{hyperref}

\definecolor{darkred}{rgb}{0.6,0.0,0.0}
\definecolor{darkgreen}{rgb}{0.0,0.6,0.0}

\usepackage{mathrsfs}
\newcommand{\ve}[1]{\textbf{#1}}		
\DeclareMathOperator{\E}{\mathbb{E}}
\DeclareMathOperator*{\argmin}{arg\,min}

\begin{document}

\title{Masked Generative Video-to-Audio Transformers with Enhanced Synchronicity } 

\titlerunning{MaskVAT with Enhanced Synchronicity}

\author{Santiago Pascual\inst{1} \and
Chunghsin Yeh\inst{1}\thanks{Corresponding author: \small\texttt{cyeh@dolby.com}} \and
Ioannis Tsiamas\inst{1,2} \and
Joan Serr\`a\inst{1}
}

\authorrunning{S.~Pascual et al.}


\institute{Dolby Laboratories \and
Universitat Polit\`ecnica de Catalunya\\
}

\maketitle

\begin{abstract}
  Video-to-audio (V2A) generation leverages visual-only video features to render plausible sounds that match the scene. Importantly, the generated sound onsets should match the visual actions that are aligned with them, otherwise unnatural synchronization artifacts arise. Recent works have explored the progression of conditioning sound generators on still images and then video features, 
  focusing on quality and semantic matching while ignoring synchronization, or by sacrificing some amount of quality to focus on improving synchronization only. 
  In this work, we propose a V2A generative model, named MaskVAT, that interconnects a full-band high-quality general audio codec with a sequence-to-sequence masked generative model. This combination allows modeling both high audio quality, semantic matching, and temporal synchronicity at the same time. 
  Our results show that, by combining a high-quality codec with the proper pre-trained audio-visual features and a sequence-to-sequence parallel structure, we are able to yield highly synchronized results on one hand, whilst being competitive with the state of the art of non-codec generative audio models. Sample videos and generated audios are available at \url{https://maskvat.github.io/}.
\end{abstract}

\section{Introduction}
\label{sec:intro}

Audio-visual cross-modal generation has gained a lot of traction in recent years, with the appearance of works for both audio-to-video (A2V) and video-to-audio (V2A) generation~\cite{chen2017deep, hao2018cmcgan,zhou2018visual,chen2020generating,dong2023clipsonic,wang2023v2a,jeong2023power,luo2024diff}. 
V2A generation has some immediate and impactful applications for the media production industry. On the one hand, it promises to accelerate, improve, and/or simplify foley sound effect generation. 
On the other hand, tasks that feature both synchronization with respect to a visual input and also a textually guided conditioning, 
like automatic dubbing, can greatly benefit from a synchronized V2A generative model that features multi-modal conditioning.

Autoregressive (AR) and mask-based deep generative models operate on discrete latent spaces. These generative strategies have been repeatedly applied to audio generation tasks recently~\cite{borsos2023soundstorm,copet2024simple,agostinelli2023musiclm,borsos2023audiolm, mei2023foleygen, kreuk2022audiogen}, thanks to innovations coming from the neural audio codec field~\cite{defossez2022high,kumar2024high}. Therefore, the utility of audio codecs is being partially re-purposed as generation facilitators, 
turning any audio processing task into a language/token processing one. Especially relevant is the fact that recent neural codecs also learn from a large variety of sound types~\cite{defossez2022high}, 
and some even compress full-bandwidth (44.1\,kHz sampling rate) general sounds 
into low bit-rate (e.g.,~8~kbps) token streams~\cite{kumar2024high}.

In this work, we propose the Masked Generative Video-to-Audio Transformer (MaskVAT), a V2A system that interconnects a state of the art full-band general audio codec with a masked generative modeling approach, bridging them with a variety of multi-modal audio-visual features that drive the V2A generation. We investigate the effectiveness of these driving features from three different performance angles. Firstly, we aim to maximize the generated audio quality by leveraging the full-band general audio codec. Secondly, inspired by the effectiveness of previous V2A works in bridging pre-trained foundation models~\cite{wang2023v2a}, we tackle the semantic matching in a similar fashion. Thirdly, we focus on the temporal alignment problem of the generated audio with respect to the input video with special emphasis. This objective is realized by employing a sequence-to-sequence model architecture, incorporating a regularization loss to ensure video-audio synchronization during generation, using a set of pre-trained synchronicity features, and implementing a post-sampling selection model.



\section{Related Work}
\label{sec:related_work}

\subsection{Video to Audio Generation} 
Early neural V2A approaches proposed sound synthesis from videos as a way to study physical interactions of materials within a visual scene of limited diversity~\cite{owens2016visually}. Similarly, other early works started tackling V2A inside a cross-modality generative adversarial framework, where both V2A and A2V were tackled as a joint problem~\cite{chen2017deep, hao2018cmcgan}. 
A number of source-specific models (targeting specific video/sound classes) were also proposed~\cite{zhou2018visual, chen2020generating}.


Motivated by the need to scale V2A as a source-agnostic problem, SpecVQGAN~\cite{iashin2021taming} was proposed as a first multi-class visually-guided sound generator model. SpecVQGAN is built upon an autoregressive transformer~\cite{vaswani2017attention} that learns to generate sequences of codewords that represent mel spectrograms through a VQGAN lossy compression~\cite{esser2021taming}. Then, a neural vocoder is used to invert the mel spectrogram back into the audio waveform.
Im2Wav~\cite{radford2021learning} is another Transformer-based audio language model conditioned on image representation to perform V2A. In this case, a pre-trained CLIP model is used to extract the sequence of visual features coming from the video frames. Then, their approach predicts the discrete tokens obtained from a VQ-VAE model~\cite{razavi2019generating}. Similarly, CLIPSonic-IQ~\cite{dong2023clipsonic}  leverages the CLIP features of individual visual frames to drive a sound generator. In this case, their generative approach follows a diffusion strategy that generates mel spectrograms. This skips the usage of a lossy compression, but requires a neural vocoder to produce audio waveforms, like many previous works. 
Except for the usage of a pre-trained CLIP encoder, all these proposals train multiple modules from scratch with their own limited data collections. 

The potential of multiple prior-mapping models 
has been recently investigated. 
In particular, V2A-mapper~\cite{wang2023v2a} bridges the domain gap between an average CLIP embedding, which summarizes the input video sequence, and a CLAP embedding, which drives an AudioLDM generative model.
Many works leverage visual encoders that were pre-trained for individual image recognition tasks~\cite{iashin2021taming, sheffer2023hear, dong2023clipsonic, wang2023v2a}. However, this usually hinders the process of modeling the visual dynamics intrinsic to the video scene and its audio-visual synchronicity. 
Diff-Foley~\cite{luo2024diff} was proposed to improve this, by developing latent-diffusion generative model which is driven by a contrastive audio-visual pre-trained (CAVP) encoder. The CAVP explicitly learns to distill audio onset features into the video encoder through self-supervised training, fine-tuning a video encoder to extract alignment-sensitive visual cues  to drive the V2A generation. On similar line, FoleyGen~\cite{mei2023foleygen} proposes specific architectural attention patterns pre-designed to enforce audio-visual alignment in their generative model. 


\subsection{Audio-Visual Alignment Representations}
\label{sec:related_av_alignment}
A crucial aspect of V2A is the synchronization (temporal alignment) between an input video and the generated audio. This is often achieved with the help of an audio-visual alignment representation model. AVST (Audio-Visual Synchronisation with Transformers)~\cite{chen2021audio} detects audio-visual synchronisation in a self-supervised manner and predicts the class as either sync or off-sync. SparseSync~\cite{iashin2022sparse} considers that the audio-visual correspondence may only be available at sparse events. The proposed SparseSelector compresses the audio and visual input tokens into two small sets of learnable selectors. These selectors form an input to a transformer which predicts the temporal offset between the audio and visual streams. It formulates audio-visual synchronisation as a classification task onto a set of offsets (for example, 21 classes between $-2$/$+2$\,sec.). As mentioned, Diff-Foley~\cite{luo2024diff} adopts CAVP to learn more temporally and semantically aligned features, then it trains a latent diffusion model~(LDM) with CAVP-aligned visual features on spectrogram latent space. That is, it leverages CAVP for (1) generating audio that is temporally aligned with the visual events, and (2) deriving the Alignment Accuracy metric. 

\subsection{Autoregressive and Mask-based Audio Token Generation}
\label{sec:mask_token_related_work}
Early works proved that waveform-based generative modeling was possible with explicit maximum-likelihood autoregressive (AR) strategies, as in WaveNet~\cite{oord2016wavenet} or SampleRNN~\cite{mehri2016samplernn}. These proposals suffered from inefficiencies inherent to their AR nature, which was palliated by subsequent works like WaveRNN~\cite{kalchbrenner2018efficient} or parallel WaveNet~\cite{oord2018parallel}.
The advancement in neural audio codecs also facilitated the use of language modeling strategies for generative audio, and one of their strong advantages over previous models is the lower framerate featured in the codec spaces compared to the raw waveforms. Based on the SoundStream codec~\cite{zeghidour2021soundstream}, AudioLM~\cite{borsos2023audiolm} is the first to take a language modelling approach to audio generation, which combines semantic and acoustic tokens in a hierarchical fashion to achieve long-term consistency and high quality. Based on Encodec~\cite{defossez2022high}, AudioGen~\cite{kreuk2022audiogen} is an AR generative model that generates audio samples conditioned on text inputs. Following AudioLM, MusicLM~\cite{agostinelli2023musiclm} tackles conditional music generation by means of a hierarchical sequence-to-sequence modeling approach based on MuLan audio tokens~\cite{huang2022mulan} in addition to the semantic tokens and acoustic tokens in~\cite{borsos2023audiolm}. Following AudioGen, MusicGen~\cite{copet2024simple} consists of an AR transformer-based decoder conditioned on a text or melody representation. 

Despite promising results are obtained in the aforementioned models, the AR sequence length grows quadratically, easily forming an extremely long sequence due to the temporally-dense nature of audio and the multiple levels of VQ codebooks. SoundStorm~\cite{borsos2023soundstorm} is one of the first to adapt a parallel decoding scheme like MaskGIT~\cite{chang2022maskgit} to predict masked audio tokens produced by SoundStream~\cite{zeghidour2021soundstream}. Based on DAC~\cite{kumar2024high}, VampNet~\cite{garcia2023vampnet} follows a similar approach for music audio generation. Through different prompting techniques, VampNet can operate in a continuum between compression and generation. Based on Encodec~\cite{defossez2022high}, MAGNet~\cite{ziv2024masked} proposes to further improve the efficiency and quality by means of predicting spans of masked tokens, scoring the prediction confidence with a pre-trained model, and fusing AR and non-AR generation.

\section{Method}
\label{sec:method}

\subsection{Audio Tokenizer}
\label{sec:audio_tokenizer}

In this work, we consider full-band single channel audio sequences. This means that our model has to process waveforms of audio sampled at 44.1\,kHz or more.  In order to decouple audio quality from the scalability of our generative strategy, we choose to operate in a latent space of low framerate, 
and since our strategy follows a discrete masked-token framework, our latent encoder must feature some discretized bottleneck at its core. To this end, we leverage a state of the art pre-trained neural codec for general audio, the Descript audio codec\footnote{\url{https://github.com/descriptinc/descript-audio-codec}} (DAC)~\cite{kumar2024high}. 
DAC takes an audio waveform of $T$ samples $\ve{x}^a \in \mathbb{R}^T$ and returns a codegram, which is a tensor $\ve{C}^a = \text{DAC}(\ve{x}^a)$, where $\ve{C}^a \in \mathbb{R}^{L\times K}$. A strong convenience of DAC is the framerate reduction it features, converting the waveform at 44.1\,kHz to $K$ token sequences of 86.1\,Hz. 
The number of parallel channels in $\ve{C}^a$ refers to the amount of RVQ levels, which hierarchically increase the codec bitrate while maintaining the same sequence length $L$. In our case, we stick to the pre-trained DAC with $K=9$. 

\subsection{Masked Generative Video-to-Audio Transformer}

\begin{figure}[tb]
	\centering
	\includegraphics[width=0.9\textwidth]{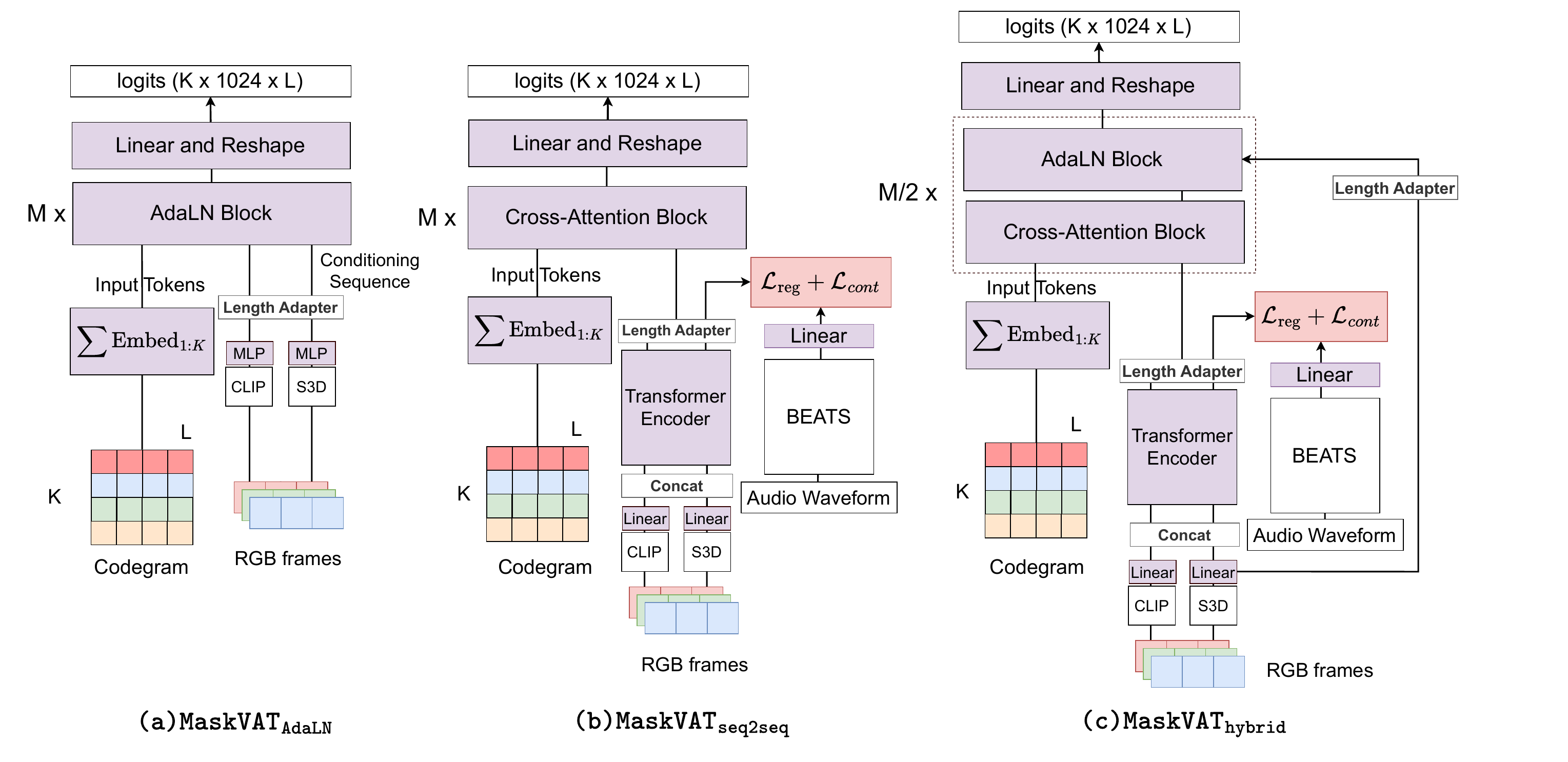}
	\caption{Overview of the three main MaskVAT structures proposed. 
 }
	\label{fig:maskgits_overview}
\end{figure}

Similarly to recent works~\cite{borsos2023soundstorm, garcia2023vampnet}, our generative strategy follows the formulation introduced in masked generative token modeling from computer vision (MaskGIT~\cite{chang2022maskgit}), which was adapted to perform masked acoustic token modeling~\cite{garcia2023vampnet}. Therefore, we have a Transformer architecture that predicts the tokenized sequence of audio. A key difference with respect to the image domain is the usage of a hierarchical tokenizer to compress the raw audio through the RVQ neural codec~\cite{garcia2023vampnet}. Considering the codegram $\ve{C}^a \in \mathbb{R}^{L\times K}$ coming from the tokenizer (see Sec.~\ref{sec:audio_tokenizer}), the output of our MaskVAT directly yields the probabilities for the whole codegram, spanning $K$ levels and $L$ time-steps, all in parallel. This is represented in the logits of the different explored models in Fig.~\ref{fig:maskgits_overview}.
Nevertheless, since the summation of RVQ levels embeddings from the audio tokenizer is intrinsically representing a full-band acoustic composition~\cite{kumar2024high}, we follow this strategy of first embedding and then summing the codewords in order to obtain the input tokens for our MaskVAT Transformer. Moreover, we initialize the embeddings with the pre-trained RVQ embeddings from DAC, provided that during early experiments we found beneficial to leverage them for faster convergence. This codegram embedding per $k$-level and summation is depicted in Fig.~\ref{fig:maskgits_overview}, having as many embedding parallel layers as $K$ codegram levels coming from the audio tokenizer. 
Once we obtain the embedded input token sequence, we inject it into a Transformer model, which is built based on two main possible blocks. On the one hand, we can use an adaptation of the AdaLN block proposed in diffusion Transformers~\cite{peebles2023scalable}. 
This modification adapts the AdaLN modulation to deal with conditioning sequences, hence featuring a temporal dimension of information. Both input tokens and conditioning sequence must feature the same length in this case. 
On the other hand, we can also consider the usage of cross-attention as a way of learning the alignment between the conditioning sequence and the audio token sequence. 

Fig.~\ref{fig:maskgits_overview} shows the diagrams of three designs explored in this work. The first one is $\text{MaskVAT}_\text{AdaLN}$, depicted in Fig.~\ref{fig:maskgits_overview}-a, which stacks $M$ AdaLN blocks to build the Transformer structure. The conditioning front-end outputs are adjusted to have the same length as the Transformer input token sequence through the length adapter, which is a nearest neighbor interpolation layer, and get concatenated channel-wise after-wards to be served to the AdaLN blocks. 
The second one, depicted in Fig.~\ref{fig:maskgits_overview}-b, features a sequence-to-sequence model that first embeds visual embeddings through a transformer encoder named $\text{MaskVAT}_\text{Seq2seq}$. Then, a stack of $M$ cross-attention blocks acts as a parallel decoder in order to mix the conditioning with the main token sequence. An advantage of this approach is also the possibility of introducing auxiliary losses that enforce a semantic/alignment proximity with respect to other audio features in an end-to-end fashion. As shown in Fig.~\ref{fig:maskgits_overview}-b, the mapping in the output of the Transformer encoder is performed upon linear projections of a pre-trained BEATs encoder~\cite{chen2022beats}, i.e. $\text{Linear}(\text{BEATs}(\ve{x}_a)) \in \mathbb{R}^{N_{beats}\times H_{trn}}$, where $\ve{x}^a$ is the audio as introduced in Sec.~\ref{sec:audio_tokenizer}, $N_{beats}$ is the length of BEATs time-patch sequence, and $H_{trn}$ the hidden size of the transformer encoder. BEATs is a state of the art self-supervised audio encoder used for large scale general audio classification~\cite{chen2022beats}, therefore a good semantic descriptor of the sequence of audio events that should be aligned with the visual ones coming from the encoder branch. 
Finally, depicted in Fig.~\ref{fig:maskgits_overview}-c, we explore a hybrid approach that mixes the two previous approaches, named $\text{MaskVAT}_\text{Hybrid}$. Here we have the end-to-end learnable component of distilling BEATs into a Transformer encoder that processes the visual features, as well as the alignment enforcement of using AdaLN blocks depending on the most alignment-sensitive features, the ones coming from S3D (see Sec.~\ref{sec:visual_conditioning}). All these MaskVAT models end with a Linear head operator that yields a 3D grid of dimensions $L\times K\times D$ representing the logits over DAC codewords, where $K=9$ and $D=1024$ due to the intrinsic configuration of DAC.

\subsubsection{Visual Conditioning}
\label{sec:visual_conditioning}

The models we propose in Fig.~\ref{fig:maskgits_overview} feature two possible conditioning front-ends, which extract video features from the RGB sequences $\ve{V}^{RGB} \in \mathbb{R}^{F\times 3\times h \times w}$ to drive the V2A mapping, where $F$, $h$, and $w$ are the number of frames, their height, and their width respectively. First, a pre-trained CLIP image encoder is used to process each video frame $\ve{v}^{RGB}_f$, projected then through a time-independent MLP into the shared dimensionality of the subsequent Transformer. The motivation to use CLIP as a video feature encoder in our case is twofold: (1) earlier works show its effectiveness in V2A already~\cite{sheffer2023hear,dong2023clipsonic,wang2023v2a}, and (2) its multi-modal nature expands the applicability of our proposal to text-driven video-editing applications.



Secondly, we also consider a 3D convolutional video encoder named S3D~\cite{xie2018rethinking}. We take the pre-trained version of S3D built in the SparseSync work\footnote{\url{https://github.com/v-iashin/SparseSync}}, for detection of audio-visual temporal offsets, i.e. detecting temporal shifts between the two modalities~\cite{iashin2022sparse}. The authors of SparseSync originally took S3D pre-trained on the Kinetics 400 dataset for video activity recognition~\cite{kay2017kinetics}, and fine-tuned S3D for the aforementioned offset detection task on AudioSet~\cite{gemmeke2017audio}. This video encoder yields a spatio-temporal tensor of features $\ve{v}_\text{S3D} \in \mathbb{R}^{N_\text{S3D} \times 512 \times h_\text{S3D} \times w_\text{S3D}}$ which we average-pool spatially to yield $\bar{\ve{v}}_\text{S3D} \in \mathbb{R}^{N_\text{S3D} \times 512}$. We consider these features to be especially sensitive to alignment, since their pre-training task required synchronizing video activity events with the appearance of audio event onsets~\cite{iashin2022sparse}. Then, following the same procedure as CLIP embeddings, an MLP projects these features into the same dimensionality of the Transformer blocks.
AdaLN blocks get a channel-wise concatenation of these feature sequences once they are resampled to have the same lengths, resulting in the visual conditioning tensor $\ve{V} = \left[ \Phi_{N_\text{CLIP}}^{N_\text{DAC}}(\ve{v}_{\text{CLIP}});  \Phi_{N_\text{S3D}}^{N_\text{DAC}}(\bar{\ve{v}}_{\text{S3D}})\right]$, where $\Phi_{N_\text{in}}^{N_\text{out}}$ is the nearest-neighbor resampling operator~(i.e.~frame repetition) between input $N_\text{in}$ and output $N_\text{out}$ length, respectively. On the other hand, the conditioning tensor for the sequence-to-sequence encoder is the channel-wise concatenation: $\ve{V} = \left[ \ve{v}_{\text{CLIP}};  \Phi_{N_\text{S3D}}^{N_\text{CLIP}}(\bar{\ve{v}}_{\text{S3D}}) \right]$, where S3D features are adjusted to CLIP's sequence length.

\subsubsection{Training Setup}
\label{sec:training}

In a masked token modeling scenario like this, we have a codegram representation $\ve{C}^a \in \mathbb{R}^{L\times K}$ (introduced in Sec.~\ref{sec:audio_tokenizer}), and a subset of these $L\times K$ tokens is masked with a special token \texttt{[MASK]}, as shown in the training section of Fig.~\ref{fig:training_sampling_selection}. The mask positions to be replaced by \texttt{[MASK]} in the codegram $M\in\{0, 1\}^{L\times K}$ is determined by a masking scheduler function. For this work, we chose the cosine scheduler due to its proven effectiveness~\cite{chang2022maskgit}, so the probability of each position to be masked is computed as $p = cos(u)$, where $u \sim U[0, \frac{\pi}{2}]$, from which we obtain $M_{l,k} = \text{Bernoulli}(p)$. Let $\ve{C}_M^a$ be the result of applying the mask $M$ to the codegram $\ve{C}^a$ and $\ve{V}$ be the collection of conditioning features in either format of the three proposed in Fig.~\ref{fig:maskgits_overview}. The training objective is to minimize the negative log-likelihood, particularly through a cross-entropy loss, for the masked positions~\cite{chang2022maskgit}: 
\begin{equation*}
	\mathcal{L}_{mask} = - \E \left[ \sum_{\forall l\in [1,L],\forall k\in [1,K], m_{l,k} = 1} \log p(c_{l, k} | \ve{C}_M^a,\ve{V})\right] .
\end{equation*}

In the sequence-to-sequence and hybrid setups of Fig~\ref{fig:maskgits_overview}-b and Fig.~\ref{fig:maskgits_overview}-c, we use a combination of a regression + contrastive loss between the visual embedding sequence after the Transformer encoder and its corresponding BEATs-projected audio embedding sequence. Regarding regression, we apply an MSE minimization between the pairs of sequences. On the contrastive side, we pre-pend a \texttt{[CLS]} token before injecting the sequence into the Transformer encoder, and select that position as the pooled embedding representative to contrast against the average projected BEATs embedding in a CLIP-like contrastive setup~\cite{radford2021learning}. The total loss to train MaskVAT then becomes:
\begin{equation*}
	\mathcal{L}_\text{maskvat-seq2seq} = \mathcal{L}_\text{mask} +\lambda_\text{reg} \mathcal{L}_{MSE} + \lambda_\text{cont} \mathcal{L}_\text{contrastive},
\end{equation*}
where $\lambda_\text{reg}$ and $\lambda_\text{cont}$ are hyper-parameters to control the loss magnitudes of the regression and contrastive regularizations respectively. Both default to $\lambda_\text{reg} = \lambda_\text{cont} = 1$ throughout the course of this work.

\subsubsection{Sampling}
\label{sec:sampling}

Once we have trained the model to perform unmasking given the $\ve{C}_M^a$ tensor, we need  a sampling scheme in order to generate new audio codegrams from an initial fully masked instance, as depicted in the sampling section of Fig.~\ref{fig:training_sampling_selection}. Following the sampling process of the original MaskGIT~\cite{chang2022maskgit}, we first determine a number of sampling steps $N_{steps}$ depending on the computational budget. Then, we begin estimating the probability distribution of each codegram position $(l, k)$ over the codewords of the $k$-th codebook at each step $n\in [1, N_{steps}]$. While computing these probabilities, we also feature classifier-free guidance upon the logits, introducing the coefficient $\gamma$~\cite{ho2022classifier,chang2023muse}. This technique is known to improve generation quality at the expense of sample diversity. Let $l^c_n = \mathcal{M}(\hat{\ve{C}}_{M,n}^a,\ve{V})$ be the output logits of our MaskVAT $\mathcal{M}$ in conditional form, and $l^u_n = \mathcal{M}(\hat{\ve{C}}_{M,n}^a)$ be the unconditional logits that only depend on the estimated and partially-masked codegram $\hat{\ve{C}}_{M,n}^a$, our guidance-weighted logits result in $l^g_n = (1 + \gamma)l^c_n - \gamma l^u_n$~\cite{chang2023muse},
where $\gamma \ge 0$. When $\gamma = 0$, this is equivalent to a regular conditional mode in our predictions. Then, for each masked position $(l, k)$ at step $n$, we sample from the multinomial distribution. With this we generate a candidate token $\hat{c}^g_{l, k,n}$ per masked position at step $n$. Then, we compute the confidence of each of these sampled tokens based on the $\log$-probability of each position $(l, k)$. Following previous works~\cite{garcia2023vampnet, besnier2023pytorch}, we introduce a diversity term $\delta$,  which is linearly annealed throughout the $N_{steps}$ as $\delta_n = \delta \cdot (1 - \frac{n + 1}{N_{steps}})$. This is used to add noise into to the confidence computation:
\begin{equation*}
	\text{confidence}(\hat{c}^g_{l, k,n}) = \log p(\hat{c}^g_{l, k,n} | \hat{\ve{C}}_{M,n}^a,\ve{V}) + \delta_n \cdot \mathcal{N},
\end{equation*}
where $\hat{c}^g_{l, k,n}$ is a token estimate after applying guidance on its logits, at sampling step $n$, and $\mathcal{N}$ is the i.i.d. noise sample drawn from Gumbel(0,1). This diversity technique has been proven to enhance the generation quality, especially when the number of $N_{steps}$ is increased~\cite{garcia2023vampnet, besnier2023pytorch}.
In what follows, we select the next \textKappa~number of tokens to mask at the next sampling iteration $n+1$ ( according to our selected mask scheduler), take the lowest \textKappa~confidence positions of our estimates, and build a new mask by placing the $\text{[MASK]}$ values in these low confidence positions. The remaining ones are kept as successfully unmasked in the estimated codegram $\hat{\ve{C}}_{M,n+1}^a$ at the $n+1$ sampling step . This whole block of operations is repeated until $n = N_{steps}$ (as shown in Fig.~\ref{fig:training_sampling_selection}), and once we get our fully-unmasked estimated codegram $\hat{\ve{C}}^a$, we run it through the DAC decoder~\cite{kumar2024high} in order to obtain our generated waveform.

\subsubsection{Beam-based selection} 
\begin{figure}[tb]
	\centering
	\includegraphics[width=0.95\textwidth]{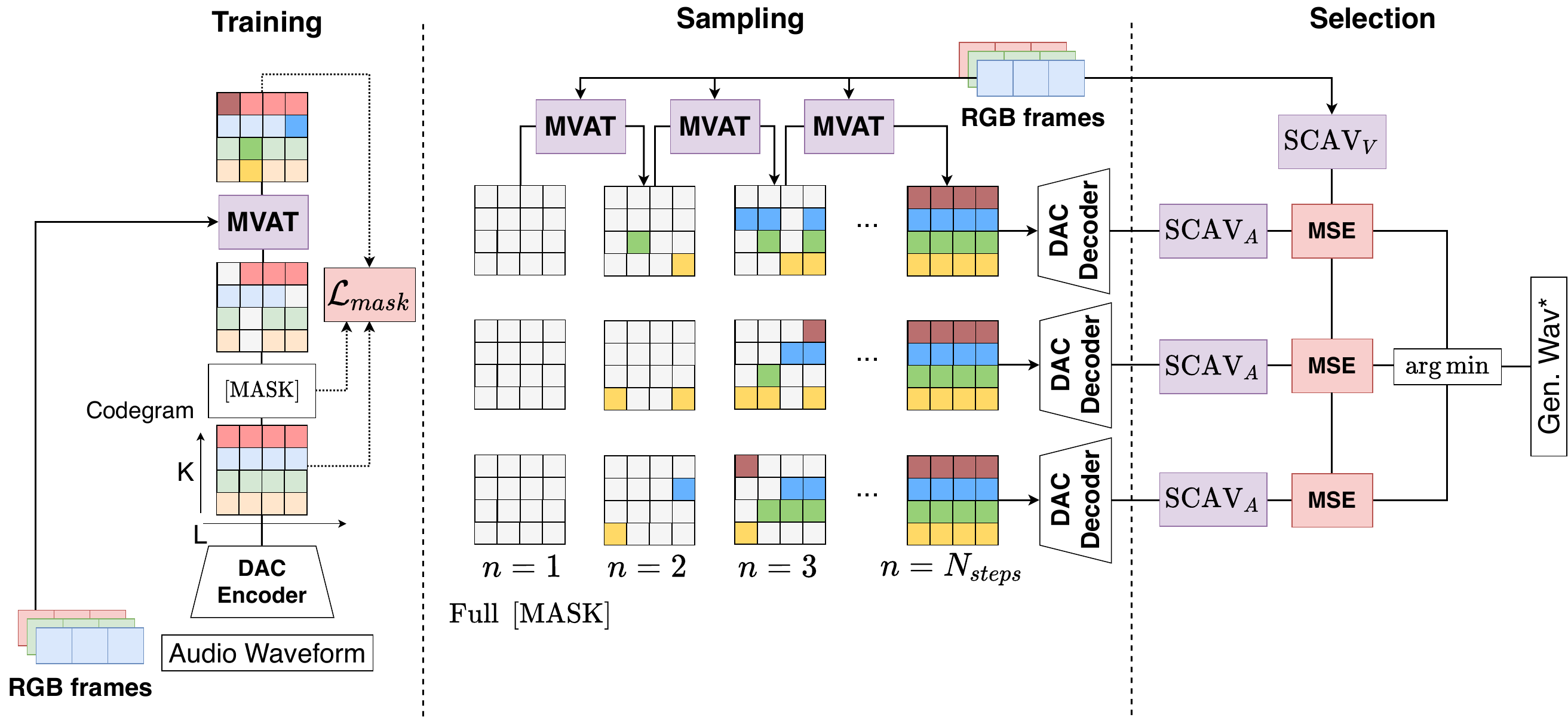}
	\caption{Overview of the Training, Sampling and Selection parts involved in the MaskVAT framework.}
        \label{fig:training_sampling_selection}
\end{figure}

The sampling process needs some tweaking of the diversity $\delta$, $N_{steps}$, and guidance $\gamma$ coefficients upon a validation set in order to produce good quality and diverse outcomes. Nonetheless, each sampling result can be very different, and some match better the input video in terms of semantic contents and alignment especially than others. In order to increase the semantic and time alignment matching with the input video, we first generate a beam-size $B$ amount of audio instances $\hat{\ve{x}}_a^i$ (exemplified with $B=3$ in Fig.~\ref{fig:training_sampling_selection}). Next, we train a sequential contrastive audio-visual~(SCAV) encoder on the same data as our MaskVAT, which maps CLIP and BEATs sequences to a common sequential space leveraging a distance-based contrastive learning approach~\cite{tsiamas2024scav}. More specifically, SCAV uses an audio and video encoder to project BEATs and CLIP features to sequences $\ve{E}^\text{scav-v}\in \mathbb{R}^{N_\text{scav}\times H_{scav}}$ and $\ve{E}_i^\text{scav-a}\in \mathbb{R}^{N_\text{scav}\times 8\times H_{scav}}$ of common length $N_\text{scav}$, and uses a contrastive loss for training that, instead of similarities between temporally-pooled sequences, leverages Euclidean distances computed between the raw sequences~\cite{tsiamas2024scav}. We use these two sequences to select the generated audio that yields the minimal distance with the input video $\hat{\ve{x}}_a^* = \argmin_i\; \text{MSE}(\ve{E}^\text{scav-v},\ve{E}_i^\text{scav-a}).$
We only use this beam strategy with $B=10$ for the subjective experiments, and provide further detail and evaluation in the Supplementary Material.

\section{Experiments}
\label{sec:experiments}
\subsubsection{Datasets}
\label{sec:datasets}
We train both our models on the VGGSound dataset~\cite{chen2020vggsound}, which contains videos curated to maximize the audio-visual correspondence in the videos while remaining unconstrained in the nature of their content. Originally, the dataset contained 200\,k video clips in their training partition, but since many videos are not available anymore and we further filter videos based on quality heuristics, we end up with a copy of approximately 155\,k video clips. The pre-processing heuristics involve removing videos with silent audio or whose audio length does not match a minimum of 10\,seconds, as well as videos featuring less than 15 video frames per second~(FPS). Each video ends up being 10\,s long, with the audio sampled at 44\,kHz, and we only use the audio-visual contents of the dataset and require no labels for the development of our work. In order to build the validation split, we selected 535 video clips re-purposed from the original train split, which amount to approximately 1.5 hours of content.

We use three test partitions to evaluate different aspects of performance for all models. First, we use a subset of the original VGGSound test split to evaluate the generated audio quality and semantic matching with the video (see Sec.~\ref{sec:obj_metrics}). In this VGGSound-test spilt we end up with 12,639 video clips after following the same process as in train split. Secondly, we assess the temporal alignment only on a subset of VGGSound-test specifically filtered to contain only sparse in time-and-space synchronisation signals~\cite{iashin2022sparse}). This subset contains videos whose audio events and their onsets exhibit strong alignments sparsely in time, like a dog barking in-camera, or a tennis player hitting the ball.  This split is named VGGSound-test-sparse. Third, we also test each model's capabilities out-of-distribution (OOD) on the music synthesis domain by leveraging the MUSIC dataset. The nature of these videos also requires strong audio onset detection from the close-up camera recording of someone playing a musical instrument, therefore we use this dataset to evaluate audio quality, semantic matching, and temporal alignment combined~\cite{zhao2018sound, zhao2019sound, dong2023clipsonic}. We extracted 1,908 test video clips, each spanning 10\,s duration, from a non-overlapped sliding window applied upon 103 test videos effectively downloaded from the MUSIC21-solo test partition~\footnote{\url{https://github.com/roudimit/MUSIC_dataset}}.

\subsubsection{Baselines}
The baselines we choose to compare against our MaskVAT variations are SpecVQGAN~\cite{iashin2021taming}, Im2Wav~\cite{sheffer2023hear}, V2A-Mapper~\cite{wang2023v2a}, and Diff-Foley~\cite{luo2024diff}, all of them introduced in Sec.~\ref{sec:related_work}. V2A-Mapper is considered a state of the art image/video-to-audio generator for VGGSound, hence we consider it our strongest competitor in quality terms. However, it does not model synchronicity explicitly. Therefore, we consider Diff-Foley as the strongest competitor in terms of alignment, since their work emphasizes a solution upon this problem. Since all baselines feature 16\,kHz audio generations, except for SpecVQGAN with 22.05\,kHz, we also run a band-width extension~(BWE) algorithm based on AudioSR~\cite{liu2023audiosr} upon them. This is done to compare fairly against our MaskVAT, which natively generates 44.1\,kHz audio and would have a trivial advantage in quality/fidelity evaluation due to a wide-band vs.\ full-band comparison.

\subsubsection{Implementation Details}

All models were trained until convergence, tracking the aggregated score of the masked token prediction accuracy, the average FD scores, and the WavCLIP score on the VGGSound validation partition. For $\text{MaskVAT}_\text{AdaLN}$ model variations, we trained with an effective batch size of 200 (across 4 GPUs). For $\text{MaskVAT}_\text{Seq2Seq}$ and $\text{MaskVAT}_\text{Hybrid}$  approaches, we trained with an effective batch size of 400 (across 8 GPUs). Larger batches helped stabilizing convergence in this case, probably due to the contrastive component upon the CLIP+S3D encoder in $\mathcal{L}_{contrastive}$ (see Sec.~\ref{sec:training}). We used AdamW~\cite{loshchilov2017decoupled}, applying a weight decay of $10^{-5}$, a learning rate warmup for the first $3,000$ iterations and polynomial decay between $10^{-6}$ and $2\cdot 10^{-4}$ for the rest. Additionally, all models were trained with 10\,\% conditioning dropout that replaced visual conditionings by learnable $\left[\text{NULL}\right]$ tokens representing the unconditional mode for classifier-free guidance~\cite{ho2022classifier} (see Sec.~\ref{sec:sampling}).

For classifier free-guidance, we explored $\gamma$ values within $[0, 16]$, and similarly, for diversity, we investigated $\delta$ within the same range. The number of sampling steps, $N_\text{steps}$, was varied from 8 to 128. As optimal settings across MaskVAT variations, we identified $N_\text{steps} = 32$, $\gamma \in \left[2, 4\right]$,and $\delta = 8$. We also found beneficial to feature the post-sampling selection with increasing beam size $B$ through  objective scans. More information about this hyper-parameter scans is available in the Supplementary Material.


\subsubsection{Objective Metrics}
\label{sec:obj_metrics}

During the development of our V2A experiments we evaluated three axes of performance: (1) generated audio quality, (2) semantic matching between the generated audio and the original audio/video, and (3) temporal alignment between the generated audio and the original audio/video.
For an objective measurement of quality, we rely on computing the Fréchet distance (FD) upon different audio feature extractors, as done by previous audio synthesis works~\cite{kilgour2018fr,pascual2023full, gui2023adapting, liu2023audioldm, wang2023v2a, dong2023clipsonic}. FD is supposed to rate the trade-off between quality and diversity attained in the generated audio. Moreover, each audio feature extractor used to compute a different FD offers a different focus on aspects of the generated audio that fit those of the ground truth~\cite{gui2023adapting}. In this work, we leverage three types of embeddings to compute FDs. First, we use VGGish~\cite{hershey2017cnn} to yield the more standardized Fréchet audio distance~(FAD~\cite{kilgour2018fr}) for better comparability with the state of the art. This is a classifier working on magnitude filter-bank representations of the audio, with a receptive field of one second, that operates on 16\,kHz signals. Secondly, we use an  MFCC representation to obtain the $\text{FDM}$ metric. This representation is frame-based, with each frame containing a window of 2048 samples and a shift of 512 samples. We extract 128 filter-banks and 64 MFCCs, so the embeddings to compute the $\text{FDM}$ are 64-dimensional. 
Finally, we also leverage the DAC codec 8-dimensional embeddings across the $K$ RVQ levels after quantization, prior to the residual summation at the input of the decoder. This is the FDD metric, and the dimensionality of the embeddings are $8\times K$, which is $8\times 9 = 72$ in the default pre-trained DAC used in this work. This is also a frame-based representation, with a wider receptive field than the MFCC one. Importantly, both MFCC and DAC front-ends operate on 44.1\,kHz signals, hence measuring the statistical distance in the full-band scenario, which is important for a general audio synthesis situation like ours~\cite{pascual2023full}. 

To assess the semantic matching, we propose two metrics that measure the proximity of the signals in the highly semantic CLIP space, as other generative works proposed~\cite{yariv2023avalign}. Nonetheless, since we generate audio to be evaluated instead of images or text/labels, we leverage the audio waveform front-end Wav2CLIP~\cite{wu2022wav2clip} to project our generated outcomes into CLIP space. Wav2CLIP was precisely trained to project 16\,kHz audio waveforms of variable length into a fixed embedding in the CLIP space from audio-video data~\cite{radford2021learning}. Then, we measure the cosine similarity between the two projected embeddings. We implement two ways of projecting through Wav2CLIP to measure the proximity: first, we project both the generated audio and the ground truth audio that came originally with the video. Then, we measure the cosine similarity of both embeddings L2 normalized in CLIP space. We name this metric WaveCLIP~(WC), where higher values imply closer semantic audio-vs-audio. As a complementary variant, we project only the generated audio and compare it against the average video CLIP embedding, both L2-normalized. We name this metric CyleCLIP~(CC), since we evaluate how aligned is the generated waveform with the original visual content.

Finally, we measure the degree of alignment of generated audios with two metrics. We compute the self-similarity-based audio novelty, reported as novelty score~(NS)~\cite{foote2000automatic}. This is obtained as the Pearson correlation coefficient between the self-similarity audio novelty curves of the BEATs-encoded sequences~\cite{chen2022beats} for the generated and ground truth audio signals. Note that video prompts are not involved in this metric, hence it is an audio-to-audio comparison.
We also consider the SparseSync~($\text{SS}$) metric, based on the synchronization model proposed in~\cite{iashin2022sparse} (see Sec.~\ref{sec:related_av_alignment}), as the mean offset prediction originally proposed in~\cite{iashin2022sparse} between the prompted video with our generated audio: ($\text{video}_{i}$,~$\text{genaudio}_{i}$). 

\subsubsection{Subjective Evaluation}


We also set up a subjective test that features three sections explicitly asking 19 human subjects (with 11 audio processing experts) to rate: (1) audio quality and relevance (as semantic matching), (2) audio-video alignment, and (3) overall quality (mix of audio quality, semantic matching, and temporal alignment). For (1), we have selected  samples from VGGSound-test containing both sparse events~\cite{iashin2022sparse} (attack sounds with distinct onsets) and dense events (sustained sounds with temporal evolution). We take into account both 16\,kHz and 44.1\,kHz versions for all models and references for comparison. This was done by running the bandwidth-extension algorithm upon the baselines, or by resampling the references or MaskVAT generation down to 16\,kHz. For (2), we have selected  samples from VGGSound-Test-Sparse containing only sparse events. In order to focus on the audio-visual synchronicity, we use samples of 16\,kHz only. For (3), we select samples from the MUSIC dataset, because the audio is highly correlated with the video and the music audio is of high quality, which is challenging to generate. Users are asked to rate (3) with all the criteria in mind (quality + semantic + alignment). Here we keep the original sample rate for all the samples such that the overall advantage of a model can be evaluated. More details about the subjective test setup can be found in the Supplementary Material.

\section{Results}
\label{sec:results}

\newcolumntype{A}{>{\raggedright}p{0.23\textwidth}}
\newcolumntype{B}{p{0.1\textwidth}}
\newcolumntype{C}{>{\centering\arraybackslash}p{0.09\textwidth}}

\begin{table}[tb]
  \caption{Objective Results on VGGSound-Test. Baselines featuring bandwidth extension to 44.1\,kHz have a BWE suffix~(e.g. V2A-Mapper-BWE). $\text{MaskVAT}_\text{AdaLN-A}$: only CLIP conditioning. $\text{MaskVAT}_\text{AdaLN-B}$: CLIP and S3D conditioning.}
  \label{tab:fd_results_vggsound}
  \centering
  \resizebox{0.95\textwidth}{!}{
  \begin{tabular}{A|CCC|CC|CC}
    \toprule
    \multirow{2}{*}{Model} & \multicolumn{3}{c}{Quality} & \multicolumn{2}{c}{Semantic} & \multicolumn{2}{c}{Alignment} \\
    \cmidrule{2-8}
    {} & $\text{FDD}\downarrow$ & $\text{FDM}\downarrow$ & $\text{FAD}\downarrow$ & $\text{WC}\uparrow$ & $\text{CC}\uparrow$ & $\text{NS}\uparrow$ & $\text{SS}\downarrow$ \\
    \midrule
    DAC reconstruct  & 0.04 & 0.10 & 1.06 & 0.90 & 0.126 & 0.97 & 0.46 \\
    \midrule
    Diff-Foley  & 1.09 & 30.8 & 8.60 & 0.35 & 0.087 & 0.07 & 0.57 \\
    Diff-Foley-BWE  & 1.22 & 22.4 & 7.54 & -- & -- & -- & -- \\
    Im2Wav  & 0.45 & 11.9 & 6.21 & 0.45 & 0.116 & 0.00 & 0.68 \\
    Im2Wav-BWE & 0.45 & 7.24 & 7.89 & -- & -- & -- & -- \\
    SpecVQGAN  & 0.26 & 7.75 & 5.27 & 0.33 & 0.080 & 0.02 & 0.67 \\
    SpecVQGAN-BWE  & 0.42 & 8.10 & 5.75 & -- & -- & -- & -- \\
    V2A-Mapper  & 0.45 & 14.1 & 0.89 & 0.47 & 0.124 & -0.01 & 0.68 \\
    V2A-Mapper-BWE & 0.24 & 2.72 & \textbf{0.84} & -- & -- & -- & -- \\
    \midrule
    $\text{MaskVAT}_\text{AdaLN-A}$ & 0.06 & 1.21 & 3.83 & 0.48 & 0.123 & 0.05  & 0.60 \\
    $\text{MaskVAT}_\text{AdaLN-B}$ & \textbf{0.05} & 0.88 & 3.39 & 0.50 & 0.123 & 0.16 & 0.43 \\
    $\text{MaskVAT}_\text{Seq2Seq}$ & 0.06 & \textbf{0.60} & 1.51 & \textbf{0.55} & \textbf{0.140} & 0.05 & 0.63 \\
    $\text{MaskVAT}_\text{Hybrid}$ & 0.08 & 0.88 & 2.04 & \textbf{0.55} & 0.136 & \textbf{0.17} & \textbf{0.40} \\
    
  \bottomrule
  \end{tabular}
  }
\end{table}

\begin{table}[h!]
	\caption{Objective Results on MUSIC-Test. Same naming conventions apply as in VGGSound-Test Results.}
	\label{tab:fd_results_music}
	\centering
    \resizebox{0.95\textwidth}{!}{
	\begin{tabular}{A|CCC|CC|CC}
		\toprule
		\multirow{2}{*}{Model} & \multicolumn{3}{c}{Quality} & \multicolumn{2}{c}{Semantic} & \multicolumn{2}{c}{Alignment} \\
		\cmidrule{2-8}
		{} & $\text{FDD}\downarrow$ & $\text{FDM}\downarrow$ & $\text{FAD}\downarrow$ & $\text{WC}\uparrow$ & $\text{CC}\uparrow$ & $\text{NS}\uparrow$ & $\text{SS}\downarrow$\\
		\midrule
		DAC reconstruct  & 0.03 & 0.17 & 7.99 & 0.88 & 0.131 & 0.94 & 0.63 \\
		\midrule
		Diff-Foley  & 0.63 & 24.2 & 46.3 & 0.43 & 0.09 & 0.02 & 0.66 \\
		Diff-Foley-BWE  & 0.52 & 22.5 & 47.7 & -- & -- & -- & -- \\
		Im2Wav  & 0.49 & 14.1 & 38.4 & 0.38 & 0.08 & 0.00 & 0.69 \\
		Im2Wav-BWE & 0.49 & 6.63 & 44.7 & -- & -- & -- & -- \\
		SpecVQGAN  &0.27 & 7.06 & 43.2 & 0.29 & 0.07 & 0.01 & 0.68 \\
		SpecVQGAN-BWE  & 0.41 & 7.18 & 44.5  & -- & -- & --  & -- \\
		V2A-Mapper  & 0.55 & 14.4 & 12.8 & 0.56 & 0.124 & 0.01 & 0.68 \\
		V2A-Mapper-BWE & 0.30 & 4.81 & \textbf{12.1} & -- & -- & -- & -- \\
		\midrule
		$\text{MaskVAT}_\text{AdaLN-A}$ & 0.08 & 1.60 & 22.8 & 0.53 & 0.123 & 0.02 & 0.67 \\
		$\text{MaskVAT}_\text{AdaLN-B}$ & \textbf{0.07} & 1.15 & 25.3 & 0.57 & 0.123 & \textbf{0.16} & \textbf{0.61}  \\
		$\text{MaskVAT}_\text{Seq2Seq}$ & \textbf{0.07} & \textbf{1.02} & 15.8 & \textbf{0.63} & \textbf{0.137} & 0.06 & 0.66 \\
		$\text{MaskVAT}_\text{Hybrid}$ & 0.09& 1.23 & 19.7 & 0.62 & 0.135 & \textbf{0.16} & 0.62 \\
		
		\bottomrule
	\end{tabular}
    }
\end{table}
\newcolumntype{D}{>{\centering\arraybackslash}p{0.15\textwidth}}
\newcolumntype{E}{>{\centering\arraybackslash}p{0.2\textwidth}}

\begin{table}[t]
	\caption{Mean opinion score~\cite{Taubert_mean-opinion-score_2023} results on VGGSound, $\text{VGGSound}_\text{sparse}$, and MUSIC test sets considering all subjects (top) and only audio processing experts (bottom, \dag).}
	\label{tab:subj_results}
	\centering
    \resizebox{0.95\textwidth}{!}{
	\begin{tabular}{A|DD|E|E}
		\toprule
		\multirow{2}{*}{Model} & \multicolumn{2}{c}{VGGSound} & \multicolumn{1}{c}{$\text{VGGSound}_\text{sparse}$} & \multicolumn{1}{c}{MUSIC} \\
		\cmidrule{2-5}
		{} & $\text{Fidelity}\uparrow$ & $\text{Relevance}\uparrow$ & $\text{Alignment}\uparrow$ & $\text{Overall}\uparrow$ \\
            \midrule
            Reference  & 4.09 $\pm$ 0.17 & 4.21 $\pm$ 0.18 & 4.82 $\pm$ 0.08 & 4.3 $\pm$ 0.21 \\
		Diff-Foley  & 1.58 $\pm$ 0.16 & 2.43 $\pm$ 0.21 & 2.14 $\pm$ 0.22 & 1.49 $\pm$ 0.19 \\
		V2A-Mapper  & \textbf{3.00} $\pm$ \textbf{0.19} & \textbf{3.50} $\pm$ \textbf{0.20} & 2.12 $\pm$ 0.20 & 2.44 $\pm$ 0.26 \\
		$\text{MaskVAT}_\text{Hybrid}$ & 2.76 $\pm$ 0.20 & 3.33 $\pm$ 0.21 & \textbf{3.81} $\pm$ \textbf{0.21} & \textbf{2.91} $\pm$ \textbf{0.27} \\
		\midrule
            Reference\dag  & 3.97 $\pm$ 0.23 & 4.09 $\pm$ 0.26 & 4.82 $\pm$ 0.10 & 4.15 $\pm$ 0.30 \\
		Diff-Foley\dag  & 1.47 $\pm$ 0.19 & 2.41 $\pm$ 0.28 & 2.15 $\pm$ 0.29 & 1.48 $\pm$ 0.24 \\
		V2A-Mapper\dag  & \textbf{2.92} $\pm$ \textbf{0.22} & \textbf{3.41} $\pm$ \textbf{0.25} & 2.13 $\pm$ 0.23 & 2.27 $\pm$ 0.31 \\
		$\text{MaskVAT}_\text{Hybrid}$\dag & 2.88 $\pm$ 0.25 & 3.39 $\pm$ 0.27 & \textbf{3.76} $\pm$ \textbf{0.28} & \textbf{2.91} $\pm$ \textbf{0.33} \\
		\bottomrule
	\end{tabular}
 }
\end{table}



Tables \ref{tab:fd_results_vggsound} and \ref{tab:fd_results_music} show the results evaluated with objective metrics for VGGSound and MUSIC test sets, respectively. Our proposed models beat all the baselines in FD terms across the full-band front-ends (FDD, FDM), exhibiting and advantage in natively modeling 44.1\,kHz upon DAC. Nevertheless, when comparing the more prominent low-band content with the FAD metric, MaskVAT falls behind V2A-mapper. This may imply that the usage of a lossy codec is a quality upper bound (this is clear from Table \ref{tab:fd_results_vggsound}, where the DAC reconstruction is already worse than V2A in FAD). 
Since SpecVQGAN generates 22.5kHz audio, it may make it advantageous in terms of bandwidth, compared to other baselines, which is reflected in the FDD and FDM scores. However, SpecVQGAN-BWE is not as advantageous, probably due to the difficulty of BWE given existing audio artifacts in the generated low-band content. In semantic terms, our $\text{MaskVAT}_\text{Seq2Seq}$ and $\text{MaskVAT}_\text{Hybrid}$ win over the baselines with quite a margin, indicating strong alignment of generated audio with respect to the input video, potentially due to the learnable intermediate features of the auxiliary losses. For alignment, our model wins when leveraging S3D features injected into AdaLN blocks in the architecture ($\text{MaskVAT}_\text{AdaLN}$ and $\text{MaskVAT}_\text{Hybrid}$). 

Table \ref{tab:subj_results} shows the subjective evaluation results. We see that MaskVAT outperforms all other models in the specialized categories of Alignment and Overall, and that it is still competitive with V2A-Mapper in Fidelity and Relevance. Interestingly, the gap in the latter two categories considerably shrinks when only expert listeners are considered. That is not the case with the Alignment and Overall categories, where MaskVAT remains a clear winner. Of special mention is the Alignment category, which highlights the benefit of the proposed approach for synchronicity. Another interesting thing to note is that expert listeners provided a rather low rating for the Fidelity and Relevance of the VGGSound data (bottom Reference scores), which questions the suitability of this data set to evaluate audio quality and stresses the result obtained by MaskVAT in the Overall MUSIC judgment.
\section{Conclusion}
\label{sec:conclusion}

In this work we proposed a masked generative video-to-audio Transformer, a model that generates audio based on an input silent video. MaskVAT makes special emphasis on tackling temporal alignment between the generated audio and the input video. Our solution connects a state of the art full-band general audio codec to ensure high quality outcomes, with a sequence-to-sequence masked-token generative approach, which is driven by pre-trained semantic and alignment features. Moreover, we also leverage a post-sampling selection strategy that minimizes the distance between the generated audio and the source input video. Our model outperforms existing solutions, exhibiting strong temporal alignment in the audio generations, which are fundamental in the overal resulting quality of video-to-audio generation. Furthermore, MaskVAT shows competitive performance in terms of generated audio quality and semantic relevance against previously proposed systems that leverage the inter-connection of strong foundational models to perform V2A. 



%
%
\bibliographystyle{splncs04}
\bibliography{main}

\clearpage

\end{document}